\documentclass[twocolumn,preprintnumbers,amsmath,amssymb]{revtex4}

\usepackage{graphicx}
\usepackage{dcolumn}
\usepackage{bm}
\usepackage{color} 

\begin{document}

\title{A new method for producing ultracold molecular ions}

\author{Eric R. Hudson}
\affiliation{Department of Physics and Astronomy, University of California, Los Angeles, 475 Portola Ave, Los Angeles, CA 90095, USA}%

\begin{abstract}
We propose a new method for the production of ultracold molecular ions. This method utilizes sympathetic cooling due to the strong collisions between appropriately chosen molecular ions and laser-cooled neutral atoms to realize ultracold, internal ground-state molecular ions. In contrast to other experiments producing cold molecular ions, our proposed method efficiently cools both the internal and external molecular ion degrees of freedom. The availability of an ultracold, absolute ground-state sample of molecular ions would have broad impact to fields as diverse as quantum chemistry, astrophysics, and fundamental physics; and may lead to the development of a robust, scalable quantum computer.
\end{abstract}
\maketitle

\section{Introduction}
To date, the complex internal structure of molecules has precluded the application of laser cooling, the work-horse of atomic physics, as a means to produce cold molecular samples. Thus, several techniques have been developed to provide cold molecules. The most promising of these techniques are Stark deceleration \cite{bethlem_decelerating_1999}, buffer gas cooling \cite{doyle_buffer-gas_1995}, and photo-association of ultracold atoms \cite{sage_optical_2005}. Despite enormous efforts on these fronts, success has been limited. The direct cooling methods -- Stark deceleration and buffer gas cooling -- are very general techniques that produce cold molecular samples from a large class of molecules, but their usefulness is hampered by the relatively warm temperature ($\sim$100 mK) and low density ($\sim$10$^8$ cm$^{-3}$) they produce. Likewise, because photo-associated molecules are necessarily comprised of atoms that can be laser-cooled, photo-association is not amenable to many of the interesting precision measurement and quantum chemistry experiments.

An exciting alternative to these methods is the study of cold molecular ions. Because molecular ions are easily trapped for many minutes, there are possibilities for cooling and interrogation that are not applicable to neutral molecules. Furthermore, because ion trapping is essentially species independent, many of the major goals of cold polar molecule physics can be realized with ions. The most noteworthy of these goals include: the study of cold chemistry, which not only has important implications for understanding the formation of interstellar clouds \cite{smith_ion_1992}, but will allow for investigation, and possible control, of reactive collisions in the truly quantum regime \cite{hudson_production_2006}; precision measurement of molecular transitions, which can be used to very sensitively measure parity violating effects \cite{gottselig_mode-selective_2004}, as well as constrain the possible variation of the fundamental constants \cite{flambaum_enhanced_2007}; and, perhaps most interestingly, the implementation of a scalable, quantum computation architecture \cite{andre_coherent_2006}. Molecular ions possess many of the advantages that molecules have for quantum computation with the added benefit of simple, reliable trapping. A recent proposal \cite{andre_coherent_2006} has shown how neutral polar molecules trapped near the surface of a superconducting microwave strip-line cavity would realize an excellent, scalable quantum computation architecture. This proposed quantum computation scheme relies on the coupling of the molecular rotation to the electromagnetic field of a superconducting strip-line resonator. If strong coupling can be achieved between the molecule and the field, then it was demonstrated that both the implementation of the necessary gates for quantum computation and the construction of quantum memory \cite{rabl_hybrid_2006-1} are feasible for this scalable molecule-mesoscopic system. The attractive quantum computation scheme of Ref. \cite{andre_coherent_2006} may also be applicable to cold molecular ions \cite{DemillePrivCom}, which can be easily produced at the requisite density, as long as the strong-coupling regime can be reached.

To date, cooling of molecular ions has been demonstrated through sympathetic cooling collisions with both cold helium buffer gas \cite{wang_investigations_2003} and laser-cooled atomic ions \cite{mlhave_formation_2000}. Unfortunately, neither of these methods can simultaneously produce both ultracold and internal ground-state molecules - i.e. the ground state in both the internal and external degrees of freedom, a prerequisite for the aforementioned experiments. While loading of molecular ions into traditional ion traps at cryogenic temperatures has been demonstrated with a He buffer gas, this technique, which produces limited cooling of the internal molecular degrees of freedom \cite{federer_collisional_1985,tichy_vibrational_1988}, e.g. rotation and vibration, results in relatively warm molecules (T $\sim$ 4 K). Conversely, sympathetic cooling of molecular ions via co-trapped, laser-cooled atomic ions produces no internal state relaxation due to the long-range nature of the Coulomb interaction \cite{bertelsen_rotational_2006}, but quickly cools the internal degrees of freedom down to the ultracold temperatures possible with laser cooling. While it may be possible to follow He buffer gas loading with sympathetic cooling via laser-cooled atomic ions to produce ultracold, internal ground state molecular ions, the method is unattractive due to the poor efficiency of He collisions for relaxing the internal degrees of freedom. (In Ref. \cite{federer_collisional_1985,tichy_vibrational_1988} it is shown that only 1 collision in 10$^4$ to 10$^6$ collisions leads to any internal state relaxation.) We are aware of one other well-developed proposal \cite{vogelius_rotational_2004, vogelius_probabilistic_2006} that addresses the need to cool the molecular internal degrees of freedom; however, this method is comparatively very complicated to implement, as it relies on a combination of Raman lasers, spontaneous emission, and black-body radiation to producing rotational cooling, and has yet to be demonstrated.

Here we propose a new method, based on sympathetic cooling via collisions with ultracold neutral atoms that should allow simultaneous cooling of both the external and internal molecular ion degrees of freedom. To our knowledge, this method of cooling has not been previously proposed. This is likely due to the fact that the ionization potential, IP, of atoms that are amenable to laser cooling is far less than the energy released by donating that electron to the majority of positively charged molecular ions. Therefore, when most molecular ions collide with a laser-cooled neutral atom, charge exchange occurs \cite{vogelius_probabilistic_2006} and the molecular ion may become a significantly more energetic, neutral molecule. Nonetheless, as detailed in this work, many cases exist for collisions between neutral atoms and certain molecular ions where charge exchange is energetically forbidden and the neutral-ion collision can only lead to cooling of the external and internal molecular ion degrees of freedom. Ref. \cite{makarov_radiative_2003} has proposed investigating neutral atoms collisions with both atomic and molecular ions; however, in the cases considered charge exchange was possible and the relaxation of molecular internal degrees of freedom was not discussed.

In what follows, we detail the possibilities for sympathetic cooling of molecular ions via laser-cooled neutral atoms. We provide calculations of the expected cooling rate for a specific molecular ion-neutral atom combination, and give a brief description of a simple experimental design for investigating these neutral-ion collisions.

\section{Methods}
The crux of our proposed cooling method is identifying combinations of molecular ions and neutral laser-cooled atoms that, due to energy conservation, cannot undergo charge exchange during a collision.  Towards this goal, Tab. \ref{IPTable} shows a partial list of the IP of several neutral atoms and molecules. Since the energy released when a positively charged molecular ion accepts an electron is equal to the IP of the formed neutral molecule, charge exchange is energetically forbidden if the neutral atom IP is larger than the neutral molecule IP. Note that the numbers reported in Tab. \ref{IPTable} are in eV, therefore the kinetic energy of even room temperature atoms and molecules is negligible on this scale.

\begin{table*}
\begin{tabular}{l|c|r||l r|l r|l r|l r|l r}
Atom &	IP [eV]	& $\lambda$ [nm] & Molecule	& IP [eV]	& Molecule	& IP [eV]	&Molecule	& IP [eV]	& Molecule	&IP [eV]&	Molecule&	IP [eV]\\
\hline
Be	&9.3	&235	&Rb$_2$	&3.5	&Li$_2$	&5.1	&YH$_2$	&6.2	&TiS	&7.1	&NaI	&7.64\\
Mg	&7.6	&285	&Cs$_2$	&3.6	&SrF	&5.3	&HoF	&6.2	&GeCl	&7.2	&RbBr	&7.7\\
Yb	&6.3	&399	&K$_2$	&4.1	&SrCl	&5.5	&Ti$_2$	&6.3	&CsI	&7.25	&Pd$_2$	&7.7\\
Ca	&6.1	&423	&KNa	&4.4	&SrBr	&5.5	&US	    &6.3	&SiF	&7.26	&C$_6$H$_7$N	&7.7\\
Sr	&5.7	&461	&KLi	&4.6	&SrI	&5.5	&ErF	&6.3	&GeBr	&7.3	&(Aniline)	&\\
Li	&5.4	&671	&BaI	&4.7	&CaBr	&5.6	&TiO	&6.4	&SnF	&7.4	&LiD	&7.7\\
Na	&5.1	&591	&BaBr	&4.8	&UO	    &5.7	&V$_2$	&6.4	&SnBr	&7.4	&CsBr	&7.72\\
K 	&4.3	&771	&SrF	&4.9	&CaF	&5.8	&CaO	&6.5	&Mn$_2$	&7.4	&PbBr	&7.8\\
    &       &       &CeO	&4.9	&CaH	&5.9	&SnCl	&6.6	&Si$_2$	&7.4	&Bsi	&7.8\\
    &       &       &BaF	&4.9	&CaCl	&6	    &TiO	&6.8	&MgCl	&7.5	&MgF	&7.8\\
    &       &       &PrO	&4.9	&DyF	&6	    &BaO	&6.9	&PbF	&7.5	&N(CH$_3$)$_3$&7.82\\
    &       &       &Na$_2$	&4.9	&TaO	&6	    &UN	    &7	    &PbCl	&7.5	&(Trimethylamine )	&\\
    &       &       &LaO	&4.9	&TiH	&6	    &InS	&7	    &N(CH$_2$CH$_3$)$_3$	&7.5     &LiH	&7.85\\
    &       &       &BaCl	&5.0	&SrO	&6.1	&ZrH$_2$	&7	&(Triethylamine)	&	&Cu$_2$	&7.89\\
    &       &       &LiNa	&5.0	&ZrO	&6.1	&InSe	&7.1	&HfO	&7.5	&PtSi	&7.9\\
    &       &       &VO	    &5.0	&CaI	&6.1	&TiS	&7.1	&GeF	&7.5	&Ge$_2$	&7.9\\
    &       &       &NdO	&5.0	&UC	    &6.2	&RbI	&7.1	&InTe	&7.6	&ZrN	&7.9
\end{tabular}
\caption{Partial list of the ionization potential of candidate neutral atoms and molecules. The first three columns are data for atoms amenable to laser cooling. The third column, labeled $\lambda$ [nm], lists the required wavelength for laser cooling. The remaining columns list candidate molecules and their ionization potential, IP. Since the molecular IP represents the amount of energy released when the positively charged ion of that same molecule accepts an electron, charge exchange is forbidden in collisions with a neutral atom if the molecule IP is smaller than the IP of the neutral atom. We emphasize that this is only a partial list of molecules and that the number of molecular ions that can be cooled by this technique is likely much larger. Data taken from \cite{radzig_reference_1985,_nist_????,Kaledin_1998}.
\label{IPTable}}
\end{table*}

As shown in Tab.~\ref{IPTable}, there are many possibilities for sympathetic cooling of molecular ions by laser-cooled neutral atoms.  In principle, every molecular ion considered here could be sympathetically cooled by ultracold Be atoms. However, laser cooling of neutral Be has not yet been demonstrated, due to the inconvenient wavelengths and low room temperature vapor pressure. For the sake of this proposal we choose to work with Yb atoms, which represent a compromise between high ionization potential and experimental practicality. ultracold Yb atoms are routinely produced via laser-cooling with standard diode laser technology and since they are easily ionized to produce Yb$^+$, they offer an in situ method of ion trap diagnosis. All ions preceding Ti$_2$ in Tab. I are energetically forbidden to undergo charge exchange with neutral Yb and are thus amenable to our cooling method.

Any two-body chemical reactions between these ions and neutral Yb, other than the already precluded charge exchange, would necessarily involve either removing an additional electron from the molecular ion or be an exchange-type reaction. In the former case, to donate an electron from the molecular ion would require the liberation of more energy than the IP of the molecular ion. Since double ionization potentials are typically large ($\ge10$ eV) and the electron affinity of Yb is only $\sim$0.5 eV, this is forbidden for every combination known to us. In the latter case, reactions of the form AX$^+$ + B $\rightarrow$ BX$^+$ + A may occur if the dissociation energy of the resulting BX$^+$ ion is larger than the AX$^+$ ion. Molecular ion dissociation energies are not well measured, and the possibility of this reaction must be evaluated on a case by case basis; however, since the B atom has been chosen for its large ionization potential and this reaction necessarily involves removing an electron from the B atom, this reaction is usually precluded. Furthermore, by choosing the cation and ultracold atom to be the same, \textit{i.e.} AX$^+$ + A $\rightarrow$ AX$^+$ + A, the effects of this mechanism can be mitigated. Finally,  energetically allowed addition-type reactions may exist, \textit{e.g.} Ca + CaF$^+$ $\rightarrow$ Ca$_2$F$^+$ is exothermic by $\sim$2.81 eV \cite{Kaschner_Cluster_1998}, however, they require a 3-body collision to conserve both energy and momentum and thus are not a concern at the densities considered in this work.

From our survey of available molecules, we choose to work with BaCl$^+$ in this proposal for the following reasons:

\begin{itemize}
  \item BaCl$^+$ should be easily produced via ablation of a BaCl$_2$ target.
  \item BaCl$^+$ possesses a simple ro-vibronic structure.
  \item	The dissociation energy of BaCl$^+$ is 4.8 eV, while the dissociation energy of YbCl$^+$ is 4.1 eV, thus the exchange reaction is precluded \cite{Kaledin_1998}. Furthermore, the mass of BaCl$^+$ is equal to that of neutral Yb and therefore momentum exchange during the sympathetic cooling process is favorable.
\end{itemize}

We expound each of these points in turn below:

\subsection{BaCl$^+$ Production:}
We expect that laser ablation, as detailed in Ref. \cite{espana__2004} for the production of CaF$^+$ from CaF$_2$, will efficiently produce BaCl$^+$ from BaCl$_2$. (In fact, CaCl$^+$ ions produced via laser ablation have recently been trapped \cite{DaveAndIkePrivCom}.) BaCl$_2$ sputtering targets are readily available from many suppliers. In Ref. \cite{espana__2004} ablation was performed on an optical quality sample of CaF$_2$ for the investigation of optical damage via ultra-violet laser light. We envision following the work of Ref. \cite{maussang_zeeman_2005} where ablation of a crushed and pressed sample was used to improve ablation efficiency. One advantage of this technique is that Yb (or other Yb precursors, e.g. YbCl$_2$) can be added to the sample before it is pressed. Thus, Yb$^+$ ions will also be produced in the ablation process. This will make it possible to load Yb$^+$ ions into the ion trap, allowing for trap diagnosis, molecular ion-atomic ion sympathetic cooling, and the possibility of an efficient detection scheme of trapped molecular ions through fluorescence detection of the atomic ions \cite{blythe_production_2005,van_eijkelenborg_sympathetic_1999,drewsen_nondestructive_2004}.

\subsection{BaCl$^+$ Structure:}
The positively-charged alkaline earth monohalides ions (MX$^+$) have similar electronic structure to alkaline earth chalcogens, \textit{e.g.} BeO, and are an ionically-bonded complex of two closed-shell atoms (M$^{2+}$X$^-$) - this property is both the reason that they exhibit reduced reactivity compared to other molecular ions and that their neutral atom precursor have such low IPs. Their ground-state is $^1\Sigma^+$ and thus they are described by Hund's coupling case (a). By analogy to the alkaline earth chalcogens the first excited state is $^1\Sigma^+$, though depending on the details of each molecule there may exist a lower-lying $^3\Pi^+$ state. Though the alkaline earth monohalides have been studied more than most molecular ions, very little spectroscopic data is available for any member. The case of BaCl$^+$ is somewhat aided by the observation from pyrotechnics that BaCl$^+$ can be used to produce green colors in fireworks. From this observation it is likely that the BaCl$^+$ A$^1\Sigma^+_{1/2}\leftarrow$X$^1\Sigma^+_{1/2}$ transition wavelength lies within the range between 505 nm and 533 nm \cite{pyrotechnic_2008}. As will be presented in a later section, knowledge of only the BaCl$^+$ ro-vibrational structure, not the electronic structure, is necessary to our experiments. Therefore, accurate spectroscopic identification is not necessary. The BaCl$^+$ ro-vibrational spectra can be estimated from the known constants for BaCl, where the vibrational splitting is $\omega\sim$ 8.4 THz \cite{_nist_????} and the rotational constant is B $\sim$ 1.25 GHz \cite{ryzlewicz_rotational_1982}. Since ionization of an alkaline earth monohalide results in a more tightly bound molecule, these values typically increase by $\sim$10\% (from a comparison of Ref. \cite{partridge_ab_1986} and \cite{_nist_????}) for the molecular ion over the neutral species. Nonetheless, we use these values in the remainder of this work as order of magnitude estimates.

\subsection{Sympathetic Cooling:}
\subsubsection{External degrees of freedom}
\begin{figure}
\resizebox{0.95\columnwidth}{!}{
    \includegraphics{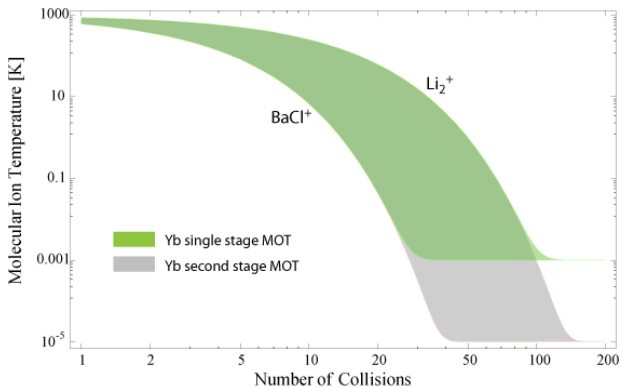}%
} \caption{(color online) Number of collisions required to thermalize the ion motion with laser-cooled Yb (m = 173 amu) atoms. The shaded regions shows the range of temperature vs. number of collisions for all of the molecular ions considered here that can be cooled via laser-cooled Yb atoms; from the best-matched BaCl$^+$ (m = 173 amu) to the worst-matched Li$_2^+$ (m = 14 amu). The green shaded region represents cooling with Yb at 1 mK, produced via cooling on the $^1$P$_1\leftarrow^1$S$_0$ line. The gray shaded region represents cooling with Yb atoms at $T$= 10 $\mu$K produced from second stage cooling on the intercombination line $^3$P$_1\leftarrow^1$S$_0$.\label{CollisionNumber}}
\end{figure}

The temperature, $T$, of a molecular ion undergoing sympathetic cooling with an ultracold Yb buffer gas at temperature $T_{Buffer}$, depends on the number of collisions, $n$, as \cite{decarvalho_buffer-gas_1999}:
\begin{equation}
T=T_{Buffer} + (T_{Initial}-T_{Buffer})e^{-n/\kappa}
\end{equation}
with $T_{Initial}$ the initial molecular ion temperature and $\kappa = \frac{(M+m)^2}{2M m}$, where $M$ is the mass of the molecular ion and $m$ is the mass of the Yb atom. A plot of Eq. 1 is shown in Fig. \ref{CollisionNumber} for the range of possible temperature versus collision number for all of the molecules considered in Tab.~\ref{IPTable} that are amenable to sympathetic cooling via ultracold Yb; from the best-matched BaCl$^+$ (m = 173 amu) to the worst-matched Li$_2^+$ (m = 14 amu). We have displayed two curves in this graph, one (green) for collisions with Yb atoms at 1 mK, produced via cooling on the $^1$P$_1\leftarrow^1$S$_0$ line, and the other (gray) for Yb atoms produced from second stage cooling on the $^3$P$_1\leftarrow^1$S$_0$ intercombination line at 10 $\mu$K.

In this plot, we have conservatively used $T_{Initial}$ = 1000 K for the initial temperature of the ions, which is the approximate temperature of species created via laser ablation; however, as detailed in a later section, we envision first loading the molecular ion trap with a cryogenic Ne buffer gas before sympathetically cooling the ions with the ultracold atoms. In this scheme, the initial ion temperature would be only $\sim$30 K, reducing the number of collisions required for thermalization by a factor of $\sim$2. Though the cooling is optimal for BaCl$^+$, as will be shown explicitly later in this section, the number of required collision for all considered molecular ions is experimentally feasible.

We now estimate the total elastic scattering rate as a function of temperature. The Eikonal approximation is valid when the interaction potential does not vary considerably over the scale of the de Broglie wavelength of the scattering particle. This approximation is appropriate as long as the collision energy, $E$, is much greater than the interaction strength - that is, as long as the idea of a semi-classical path is reasonable and the wavefunction can be replaced by a semi-classical wavefunction. With the application of the optical theorem, the Eikonal approximation gives the total elastic scattering cross-section as \cite{sakurai_modern_1985}:

\begin{equation}
\sigma_{Eik}= \frac{4\pi}{k} Im(f(k,k)),
\end{equation}
with the imaginary part of the forward-scattered amplitude given by:
\begin{equation}
Im(f(k,k))=k\int_0^\infty b(1-\cos(2\Delta(b))) db,
\end{equation}
where the phase of the semi-classical wave is found in terms of the impact parameter, $b$, as:
\begin{equation}
\Delta(b)= -\frac{\mu}{2k\hbar^2}\int_{-\infty}^\infty V(\sqrt{b^2+ z^2}) dz.
\end{equation}
In general, a molecular ion produces a polarization in a nearby neutral atom, leading to an interaction potential:
\begin{equation}
V(r)= -\frac{1}{2}\alpha|E|^2= -\frac{\alpha e^2}{2(4\pi\epsilon_o)^2 r^4}=-\frac{C_4}{r^4}
\end{equation}
where $\alpha$ is the atomic polarizability, $e$ is the absolute value of the electron's charge, and $r$ is the atom-ion separation. With straight-forward integration of Eqs. (3) and (4), we find the total elastic cross-section, as predicted by the Eikonal approximation, to be
\begin{equation}
\sigma_{Eik}=\Gamma\left(\frac{1}{3}\right)\left(\frac{\pi C_4}{2\hbar v}\right)^{2/3}\label{EikonalSigma}
\end{equation}
where $\Gamma$ is the gamma function \cite{arfken_mathematical_2001} and $v$ can be approximated by the molecular ion velocity. Finally, we have the total elastic scattering rate constant, $K$, as:
\begin{equation}
K= \left<\sigma v\right> = \Gamma\left(\frac{1}{3}\right)\left(\frac{\pi C_4}{2\hbar}\right)^{2/3}\left(\frac{8k_B T}{\pi \mu}\right)^{1/6},
\end{equation}
where $\left< \right>$ denotes thermal averaging and $\mu$ is the reduced mass. The weak T$^{1/6}$ dependence of the scattering rate constant means that sympathetic cooling is efficient even to the lowest temperatures.

\begin{figure}
\resizebox{0.95\columnwidth}{!}{
    \includegraphics{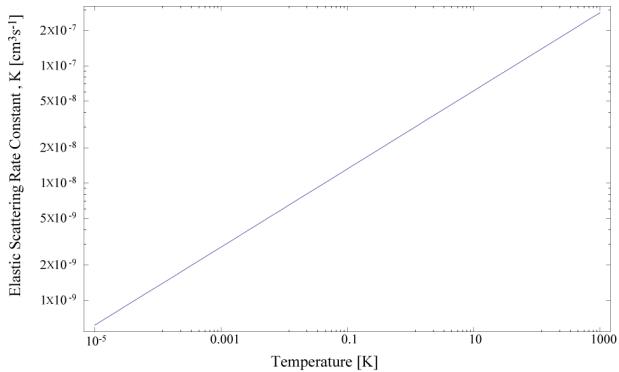}%
} \caption{Log-log plot of the elastic scattering rate constant for BaCl$^+$ + Yb collisions as given by the Eikonal approximation.\label{ElasticScatteringRate}}
\end{figure}

A log-log plot of the scattering rate constant is shown in Fig. \ref{ElasticScatteringRate} for the case of BaCl$^+$ colliding with Yb, where   $\mu$~=~173/2 amu and $\alpha=(4\pi\epsilon_o) 179$ a$_o^3$ \cite{dzuba_atomic_2007}, for the range of temperatures of concern to our project. From this figure we see that the scattering rate constant is very large by atomic physics standards (a result of the relative strength of the ion monopole field) and for typical ultracold atomic densities of $\rho$ $\sim10^{11}$ cm$^{-3}$ one would expect total scattering rates of $\gamma$ = 100 Hz - 100 kHz. Thus, the external degrees of freedom of the molecular ion are conservatively expected to sympathetically cool to $\sim$100 $\mu$K within only a few hundred milliseconds. Since molecular ions can be trapped for hours, this would be a very effective cooling mechanism. Furthermore, the vibrational state energies of typical ion traps are on the order of 10 MHz \cite{monroe_resolved-sideband_1995}, thus the temperatures represented by this cooling mechanism correspond to the molecular ions primarily occupying the absolute ground state of the external trapping potential.

An estimate of the range of validity of the Eikonal approximation can be made by defining the length scale of the interaction potential, $r_{Char}$, as equal to the value of $r$ where the ion-neutral interaction is equal to the incoming kinetic energy:
\begin{equation}
\frac{3}{2}k_BT = \frac{C_4}{r_{Char}^4}
\end{equation}
The requirement that the de Broglie wavelength be less than this characteristic length scale, leads to a constraint on the temperature for the validity of the Eikonal approximation as:
\begin{equation}
T > \frac{4\pi^4\hbar^4}{C_4 k_B \mu^2}\approx 300 \mu\rm{K}.
\end{equation}
Thus, the estimates made in this section should be accurate over the majority of the range of interest.

\subsubsection{Internal degrees of freedom}
As aforementioned, neither collisions with a helium buffer gas nor collisions with laser-cooled atomic ions are efficient at cooling the internal degrees of freedom of a molecular ion. In the former case, the inefficiency is due to the small electric polarizability of the He atom, which prevents it from coupling the molecular ion ro-vibrational states \cite{federer_collisional_1985,tichy_vibrational_1988}; in the latter case, the Coulomb repulsion between the molecular and atomic ions results in long-range collisions that do not couple the molecular ion ro-vibrational states \cite{bertelsen_rotational_2006}. Neutral atomic Yb can scatter at close range from a molecular ion and possesses a polarizability $\sim$200 times larger than that of atomic He. Since the interaction potential is linear in the polarizability, from Fermi's Golden Rule we expect that Yb may be $\sim10^4$ times more efficient than He at quenching the molecular ion internal degrees of freedom. Given that data for collisions of molecular ions with He shows that only 1 in 10$^4$ to 10$^6$ collisions relax the molecular internal states, while more polarizable buffer gases lead to higher quenching rates \cite{federer_collisional_1985,tichy_vibrational_1988}, we expect that almost every short-ranged collision with Yb should result in some relaxation of the molecular internal state.

We now calculate the expected short-range (inelastic) collision rate between BaCl$^+$ and Yb. For two colliding particles, the energy-dependent scattering cross-section for the $\ell$th partial wave, with projection $m$, from state $i$ to state $f$ in all outgoing partial waves $\ell'$, $m'$ is generally written as:
\begin{equation}
\sigma_{\ell,m}(E,i\rightarrow f)=\frac{\pi}{k^2}\sum_{\ell',m'}|T_{\ell,m,\ell',m'}(E,i\rightarrow f) |^2.
\label{SigmaLM}
\end{equation}
Here $T_{\ell,m,\ell',m'}(E,i\rightarrow f)$ is the `T-matrix', whose elements represent the probability amplitude for a transition from the incoming spherical wave, $\Psi_{i,\ell,m}$, to the out-going wave, $\Psi_{f,\ell',m'}$, and $k = \sqrt{\frac{2 \mu E}{\hbar^2}}$ is the magnitude of the wave-vector at collision energy $E$. Since we are concerned with inelastic collisions we must sum over all final states $f$, in addition to the normal sum over $\ell,m$. Furthermore, if we assume that any collision that penetrates to short-range is inelastic with unit probability, we can rewrite Eq. \ref{SigmaLM} as \cite{orzel_spin_1999}:
\begin{equation}
\sigma(E,i)  = \sum_{f,\ell,m}\sigma_{\ell,m}(E,i\rightarrow f)= \sum_\ell \frac{\pi}{k^2}(2l+1)P_T(E,\ell),
\end{equation}
where $P_T(E,\ell)$ is simply the probability of transmission to short-range.

\begin{figure}
\resizebox{0.95\columnwidth}{!}{
    \includegraphics{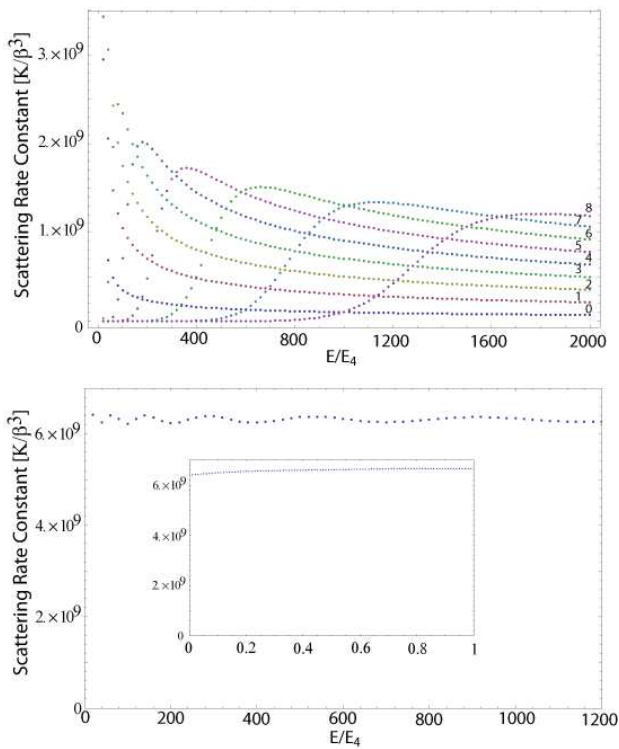}%
} \caption{Universal scattering rate constant ($K/\beta^3$) vs. Energy ($E/E_4$). Upper panels shows individual partial wave contributions, labeled by its respective $\ell$ value, for up to $\ell=8$.  The lower panel shows the total scattering rate constant, with the inset depicting the observance of the Wigner threshold scaling law for low temperature inelastic collisions. The scattering rate for a specific pair of colliding ions and atoms can be found by calculating the $\beta$ parameter, which depends only on the reduced mass and atomic polarizability. \label{InelasticScatteringRate}}
\end{figure}

Since the molecular ion-neutral atom interaction is spherically symmetric (neglecting the much weaker dipole terms) we have the 1-D Schrödinger equation as:
\begin{equation}
-\frac{\hbar^2}{2\mu}\frac{d^2\Psi}{dr^2} + \left( \frac{\hbar^2\ell(\ell+1)}{2\mu r^2} -\frac{C_4}{r^4}\right)\Psi = E\Psi.
\end{equation}
Rescaling the problem with $x=r/\beta$, where $\beta = \sqrt{2\mu C_4}/\hbar$ , and defining $E_4=\hbar^2/(2\mu\beta^2)$ we have:
\begin{equation}
-\frac{d^2\Psi}{dr^2} + \left( \frac{\ell(\ell+1)}{x^2} -\frac{1}{x^4}\right)\Psi = \frac{E}{E_4}\Psi.
\end{equation}
In this dimensionless form Eq. (4) represents the `universal' Schrödinger equation for a particle with dimensionless energy $E/E_4$ interacting with an $x^{-4}$ potential. $P_T(E,\ell)$ is calculated by numerically solving the Schrödinger equation for the potential and assuming that any flux that does not reflect from the potential is transmitted to short-range and then completely lost to inelastic processes. The `universal' solution of this problem, in terms of the scaling factors, is given in Fig. 3. To ensure convergence over the calculated range we have included up to $\ell = 8$. The individual partial waves are shown in upper panel of this graph and may be identified by their values at $E/E_4 = 2000$, where the scattering rate constant is highest for $\ell = 8$ and decreases monotonically to the lowest value for $\ell = 0$. Consistent with the Wigner threshold scaling law for low temperature inelastic scattering, our calculations find $\sigma \propto 1/k$. As a result, the scattering rate constant becomes constant as the energy goes to zero, as seen in the inset on the lower panel of Fig. \ref{InelasticScatteringRate}.

As shown in Fig. 3, the inelastic scattering rate is roughly constant over the entire energy range considered and has a value of $\sim6\times10^9/\beta^3$. For BaCl$^+$ collisions with Yb, where $\mu$ = 173/2 amu and $\alpha = (4\pi\epsilon_o) 179$ a$_o^3$ \cite{dzuba_atomic_2007}, this yields a \textbf{temperature independent} scattering rate constant of $K = 4\times10^{-5}$ cm$^3$s$^{-1}$. For a typical ultracold atomic density of $\rho = 10^{11}$ cm$^{-3}$, one would then expect a total inelastic scattering rate of $\sim$100 kHz. Thus, in only a matter of milliseconds it should be possible to quench any internal excitation of the molecules into their absolute ground state. Furthermore, even if the assumption of unit efficiency internal relaxation for Yb collisions is off by orders of magnitude the scheme is still viable -- \textit{i.e.} if only 1 in 100 collisions produced relaxation $K~\approx$ 1 kHz. With these rates it may be possible to scatter multiple photons from a single molecular ion since the inelastic collisions will relax the molecular ion ro-vibrational states to the ground state, serving as an effective dark-state repumper. Interestingly, this could lead to both efficient fluorescence detection and, possibly, direct laser cooling of the molecular ions.

As a check of the model, we predict a quenching rate constant for collisions between He and BaCl$^+$ of K $\sim 10^{-14}$ to 10$^{-16}$ cm$^3$s$^{-1}$. This is consistent with values reported in the literature for similar collisions \cite{federer_collisional_1985}.

Finally, to cool a sample of diatomic molecular ions from $T\approx1000$ K to $\sim$1 mK, requires removing an amount of energy $E = 5/2N_Mk_BT$, where $N_M$ is the total number of molecular ions and $k_B$ is the Boltzmann constant. Approximating the ultracold Yb atoms as an ideal gas we expect the temperature of the Yb atoms to rise by an amount $\Delta T = 5/3N_M/N_A T$, where $N_A$ is the total number of trapped Yb atoms. For typical values of $N_M\approx10^2$ and $N_A\approx10^8$, $\Delta T\approx 1$ mK. Thus, the heat capacity of a typical ultracold atomic sample is enough to cool the molecular ions down to ultracold temperatures. Furthermore, if the estimates here are not accurate, \textit{e.g.} the ions are produced at a higher temperature, the cooling process, as described in the next section, could simply be repeated with a fresh `batch' of ultracold Yb atoms.

\section{Experimental Implementation}
\begin{figure}
\resizebox{0.9\columnwidth}{!}{
    \includegraphics{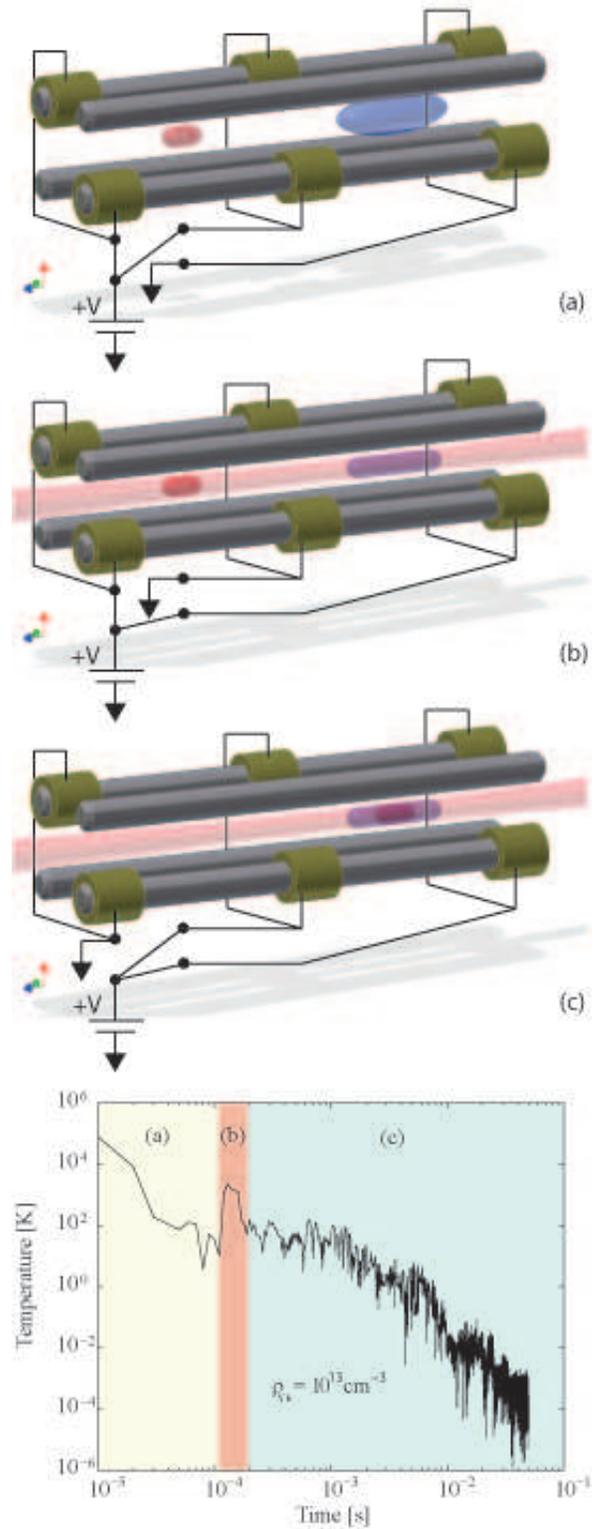}%
} \caption{(color online) (a) BaCl$^+$ ions are trapped in the left chamber, while Yb atoms are captured in a MOT centered on the right trap chamber. (b) Yb atoms are transferred into an optical dipole trap and the DC electrode voltages are switched to transfer the ions into the right chamber. (c) The DC electrode voltages are reconfigured to trap the molecular ions in the rightmost chamber, where they are overlapped with the ultracold Yb atoms. The bottom panel shows the temperature of a BaCl$^+$ ion during the three stages of the cooling process. Each shaded region, with is label (i), corresponds to the similarly labeled panel above. \label{TrapViews}}
\end{figure}

Though charge exchange is energetically forbidden for collisions between ground state ($^1$S$_0$) Yb and BaCl$^+$, Yb atoms in the excited $^1$P$_1$ state are not forbidden from donating an electron to the molecular ion. Therefore, all sympathetic cooling collisions between the neutral atoms and molecular ions must occur with the Yb atoms in their ground state. This can be arranged by using a `dual' version of the linear ion trap described in Ref. \cite{berkeland_linear_2002}. As shown in Fig.~ \ref{TrapViews}(a), this trap design consists of four linear quadrupole electrodes in the standard configuration; the radio-frequency trapping voltage is applied to the two diametrically-opposed bare electrodes, while the other two electrodes are grounded. Three pairs of short DC electrodes are slipped over the two grounded quadrupole rods. The DC electrodes are separated from the grounded rods by a thin insulator. By controlling the DC voltage applied to these `end-cap' electrodes, ions can be trapped in either the left or right `chamber', as well as shuffled between the two.

In a manner similar to that described in Ref. \cite{Hashimoto_ablation_20065} molecular ions, produced via laser ablation, can be loaded into the left chamber of the ion trap (represented by the red cloud in Fig.~\ref{TrapViews}(a)). The ion loading efficiency may be improved over that of Ref. \cite{Hashimoto_ablation_20065} by using a cryogenic Ne buffer gas \cite{antohi_cryogenic_2008} at $T\approx$ 30 K. Once the ion trap is loaded, the cryostat temperature may be lowered to $\sim$4~K, resulting in fast cryopumping of the Ne buffer gas by charcoal sorbs \cite{campbell_magnetic_2007}. Next, ultracold Yb atoms are collected in a standard MOT (represented by the blue cloud in Fig. \ref{TrapViews}(a)) and then transferred into an optical dipole trap whose axis is aligned with the ion trap axis (the blue cloud in Fig. \ref{TrapViews}(b)). Once the ultracold atoms are collected, the molecular ions are transferred from the `collection' chamber into the `collision' chamber by control of the voltage applied to the DC electrodes. By simply grounding the center DC electrode the ions will move to the right in Fig. \ref{TrapViews}. Once the ions have moved to the rightmost chamber, reapplying the DC voltage to the center electrode results in a ion trap overlapped with the ultracold atomic Yb (represented by the overlapping blue and red clouds in Fig. \ref{TrapViews}(c)). Achieving neutral atom densities of $10^{13}$ cm$^{-3}$ over the entire ion trap volume is routine with an optical dipole trap \cite{friebel_co2-laser_1998} and thus the estimates of cooling rates from the previous section are conservative.

The result of a trajectory simulation of the cooling process for a single BaCl$^+$ ion is shown in the bottom panel of Fig. \ref{TrapViews}. Here the ion `temperature' \cite{TempComment} is plotted versus time. In the first shaded region, corresponding to Fig. \ref{TrapViews}(a), the molecular ion is sympathetically cooled from the laser ablation temperature (conservatively assumed to be 10$^5$ K) down to the Ne buffer gas temperature. Approximately, 100 $\mu$s later the buffer gas is removed from the simulation and the transfer process begins (Fig. \ref{TrapViews}(b)). Initially the ion gains kinetic energy, but by timing the transfer process appropriately, it can be reconverted into potential energy. Once the transfer process is complete, the molecular ions come into equilibrium with the 10 $\mu$K Yb buffer gas. These trajectory simulations include the ion micro-motion and model the sympathetic cooling collisions as randomly directed hard-sphere collisions. For collisions with the Ne buffer gas a conservative cross-section of 2$\times10^{-14}$ cm$^2$ is used, while the Yb cross-section is calculated from Eq. \ref{EikonalSigma}.

To carefully study the BaCl$^+$/Yb collisional properties it is necessary to monitor the molecular ion external and internal temperatures. The motional temperature of the molecular ions may be measured through observation of the fluorescence from co-trapped Yb$^+$ \cite{blythe_production_2005,van_eijkelenborg_sympathetic_1999,drewsen_nondestructive_2004}. After the sympathetic cooling stage, Yb$^+$ ions can be created from ionization of the ultracold Yb atoms and the residual Yb atoms removed by extinguishing the optical dipole trap. Because of the long-range nature of the Coulomb interaction, the Yb$^+$ and BaCl$^+$ temperatures will quickly come to equilibrium. Thus, by simply measuring the atomic lineshape the molecular temperature may be inferred. This in situ thermometry will allow continual monitoring of the BaCl$^+$ temperature during the cooling process \cite{YbYbpluscomment}.

The molecular ions' internal temperature, \textit{i.e.} rotational and vibrational population distribution, can be measured using the technique of resonantly-enhanced multi-photon ionization as proposed in Ref. \cite{bertelsen_rotational_2006}. It may also be possible to measure the internal state population with a method similar to that developed for the spectroscopy of $^{27}$Al \cite{rosenband_frequency_2008}.


\section{Outlook and Summary}
Even when energetically allowed, charge exchange rates are typically much slower than elastic collisions rates \cite{makarov_radiative_2003}. Therefore, the estimates of this proposal are likely very conservative and efficient sympathetic cooling of molecular ions may be possible with more easily produced ultracold Rb or Cs atoms. Similarly, it may not be necessary to quench the ultracold atoms to their ground state, allowing continual cooling of the molecular ions by the MOT. This is especially attractive since without continued laser cooling the RF-trap induced micro-motion of the molecular ions will eventually lead to sympathetic heating of the atoms. (Though advantageous, continued cooling is not necessary. Due to the large number of atoms, this heating is negligible on the time-scales necessary to quench the ion internal motion and could easily be overcome by sympathetic cooling with a co-trapped atomic ion.)

Interestingly, our proposed technique naturally extends to a combination of an `on-chip' MOT, similar to that proposed in Ref. \cite{lewis_fabrication_2008}, and  a surface ion trap, similar to those proposed in Ref. \cite{antohi_cryogenic_2008}. Thus, using lithographic techniques it should be possible to construct simple and scalable `on-chip' ultracold molecular ion traps that will provide molecular ions in their absolute internal and external ground states, \textit{i.e.} the requisite starting point for the exciting experiments discussed in the introduction.

In summary, we have presented a new and promising mechanism for the production of ultracold molecular ions. The key realization of this work is that the commonly held belief that molecular ions cannot undergo primarily elastic collisions with atoms that are amenable to laser cooling, is not an absolute truth. We have shown that for carefully chosen ion-atom pairs, charge exchange is energetically forbidden and thus sympathetic cooling with ultracold neutral atoms is indeed possible. Therefore, this proposal represents a dramatically simplified scheme for producing ultracold molecular ions in their absolute internal and external ground state as compared to previous proposals \cite{vogelius_rotational_2004}, which rely on a four-fold combination of sympathetic cooling with atomic ions, laser induced Raman transitions, spontaneous emission, and black-body radiation. Because ion trapping is species independent, many of the major goals of cold polar molecule physics can be realized with ions. Thus, the impact of this proposal is clear. With an efficient means of producing ultracold molecular ions, interests as diverse as quantum chemistry, astrophysics, fundamental physics, and quantum computation may be pursued.

\acknowledgments
We thank D. DeMille, J.M. Hutson, B. Odom, W.G. Rellergert, and J. Ye for useful discussions.

\bibliography{Hudson_MolecularIons}

\begin{thebibliography}{47}
\bibitem[{Tem()}]{TempComment}
\bibinfo{note}{Technically, temperature cannot be defined for a single
  particle, however, for convenience we express the ion velocity as a
  temperature given by $\frac{m\vec{v}^2}{3k_B}$.}
\end{thebibliography}
\end{document}